# STATUS OF THE RFD LINAC STRUCTURE DEVELOPMENT [1]

D.A. Swenson, Linac Systems, 1208 Marigold Dr. NE, Albuquerque, NM 87122

*Abstract*

The Proof-of-Principle (POP) prototype of the Rf-Focused Drift tube (RFD) linac structure is currently under test at Linac Systems, after years of delay due to a variety of technical problems. A discussion of these technical problems and their solutions will be presented. The status of these tests will be reported. Plans for future development of this linac structure will be revealed. Potential uses of this linac structure for a variety of scientific, medical, and industrial applications will be described, including: proton linac injectors for proton synchrotrons, compact proton linacs for PET isotope production, epithermal neutron sources for the BNCT application, energy boosters for proton therapy, compact portable neutron sources for thermal neutron radiography, and pulsed cold neutron sources for cold neutron physics and related applications.

## 1 STATUS OF THE POP PROTOTYPE

The POP prototype[4,6,9] came into operation on June 19, 2000 in the Linac Systems laboratory in Waxahachie, TX. We had about 9 mA of 25-kV beam entering the RFQ and about 9 mA of beam transmitted through the RFQ and RFD linac structures. That beam impinged on a 1.25-MeV-thick absorber foil and Faraday cup assembly. The beam that passed through the absorber foil and into the Faraday cup (the accelerated beam) showed the expected threshold for both rf fields and injection energy. Because we suffer from an inadequate amount of rf power (180 kW) and are operating very close to the linac excitation threshold, the accelerated beam corresponded to only about 4% of the transmitted beam. Nevertheless, we now have conditions that produce a steady 0.3-mA beam of protons at 1.636-MeV for hours at a time.

The initial operation was sporadic as a result of several problems (in addition to the near threshold operation), which have now been rectified. We have always had a "breathing" phenomenon, which we attributed to some mechanical vibrations in the drift tube stems. A description of the solution to this problem is given below. We often witness a 10% decline in RFQ fields after 15 seconds of operation. A description of the solution to this problem is also given below.

We are confident that, with adequate rf power and the improvements that are identified below, the performance of the RFD linac structure will come up to our expectations.

## 2 BRIEF DESCRIPTION OF THE RFD LINAC STRUCTURE

The RFD linac structure[1-4] resembles a drift tube linac (DTL) with radio frequency quadrupole (RFQ) focusing incorporated into each drift tube. The RFD drift tubes comprise two separate electrodes, operating at different electrical potentials as excited by the $TM_{010}$ rf fields, each supporting two fingers pointing inwards towards the opposite end of the drift tube forming a four-finger geometry that produces an rf quadrupole field along the axis. The particles traveling along the axis traverse two distinct regions, namely, the gaps between the drift tubes where the acceleration takes place, and the regions inside the drift tubes where the rf quadrupole focusing takes place. This new structure could become the structure of choice to follow RFQ linacs in many scientific, medical, and industrial applications.

## 3 TECHNICAL PROBLEMS AND SOLUTIONS

Successful operation of the POP was delayed for more than a year by a host of minor problems, each of which have now been overcome. Descriptions of these problems and the solutions that were employed are presented below:

RFQ Alignment: The RFQ is assembled from four pieces of machined copper, namely two major pieces (top and bottom) and two minor pieces (sides), each representing one vane of the four-vane structure. The mechanical precision of these pieces was not what it should or could have been. The resulting assembly was difficult to excite in the quadrupole mode. Measuring the actual geometry and placing shims between pieces allowed the quadrupole mode to be excited. Four-rod, dipole-mode detuners, of the type developed at Los Alamos, were incorporated in the upstream end of the structure for additional stability. In the future, a tooling fixture will be used to support the major and minor pieces during machining to achieve the desired precision.

RF Power Tubes: The rf power system employed fifteen YU-141 Planar Triodes (PSI/Eimac); one in the intermediate power amplifier, IPA-1, two in IPA-2 and 12 in the final power amplifier (FPA). We started with a full complement of tubes, but quickly suffered 6 or 7 partial failures, making it impossible to reach the rated power of the system (240 kW at 600 MHz). Because of fabrication problems at Eimac, it took more than a year

---
[1] Work supported by the National Institute of Mental Health (NIMH) and the National Cancer Institute (NCI).

to get back to a full complement of good tubes and the rated power output.

Multipactoring: The twin bladed RFD stem geometry is prone to multipactor. Initially, we were unable to break through the multipactor barrier. In that situation, it is impossible to optimize the rf power system and coaxial drive line. Our first attempt to overcome the multipactoring was to coat the RFD drift tubes and stems with carbon black from an acetylene torch. This worked and allowed us to optimize the rf power system. Next, we decided to try a cleaner and more robust cure for multipactoring, namely vapor deposition of a thin layer of titanium. This also worked and allowed us to excite the RFD structure to high field levels. Some changes in the RFD stem geometry, to reduce the area of parallel surfaces, should reduce the potential for multipactor.

Cavity Q: At this point, it was obvious that the Q of the system was less than expected, resulting in a power requirement that was beyond the capabilities of the rf power system. We launched a search for the cause of the depressed Q. Several causes were under suspicion, namely the quality of the copper plating, the spring-ring joints to the end walls, the end walls, and the drift tube stems. Because of the very short tank (350 mm), the two end walls and associated rf joints have an unusually large negative effect on the cavity Q. Because of the very low injection energy (0.8 MeV), the number of drift tubes and associated stems for the short tank are unusually large, which has a negative effect on the cavity Q. In order to separate the possible causes of the depressed Q, we removed the 12 drift tubes and measured the Q of $TM_{010}$ and $TE_{111}$ modes in the empty tank. The former, which is effected by both the conductivity of the copper and the end seals, was only 38% of theoretical. The latter, which is not effected by end seal conductivity, was 68% of theoretical. We took this to mean that both the copper plating and the end seals were part of the problem.

End Seals: We decided to re-machine the tank and end walls to replace the spring ring and o-ring seals with Helicoflex copper seals, providing both rf and vacuum seals. One end of the tank was then modified to employ a custom flexed-fin rf joint backed by an o-ring vacuum seal. Future designs will employ that configuration.

Copper Plating: We decided to have the RFD tank and end plates stripped of their original copper plating and re-plated by a copper plating company that had done work for other proton linac projects. In the end, though, the cavity Q is still only 60% of what it should be.

Stem Power: A closer analysis of the RFD stem losses revealed that the stem losses were unnecessarily high and that slight changes in the geometry would reduce the stem losses by 50%. This observation, of course, was too late for the POP and we had to live with the higher stem losses. In the future, the modified stem geometry will be used.

Rf Drive Line: In a short rf drive line (approximately one wavelength) with no circulator, the length of the line has a large effect on the response of the rf system to the reflected power associated with cavity filling and cavity arcs. To provide some degree of adjustment, we installed and additional length of 3-1/8" coaxial line fitted with a sliding stub tuner. This helped to determine the optimum rf drive line length for the system.

Breathing: The RFD tank was quite sensitive to mechanical impact. Throughout this work, we were plagued with a "breathing" phenomenon, which we assumed was due to mechanical vibrations of the drift tube stems, which in turn caused a periodic oscillation in the resonant frequency and rf field levels in the structure. The driving force for this vibration appeared to be rf field, duty factor, and pulse rate dependent. This mechanical vibration was in the vicinity of 20 Hz. Variation of the pulse repetition rate in the vicinity of 20 Hz, caused the observable effects to vary significantly. Recently, we cured this effect by placing an insulating spacer between the two blades of the RFD stems at about 2/3 the distance from the outer wall to the drift tube. We plan to braze a ceramic spacer, at about that location, in all future RFD drift tubes.

Fading: We have noticed a "fading" phenomenon in the RFQ fields. We can get more fields in the RFQ after it has been off for several minutes. This field level fades by about 10% in the next 15 seconds. Recently, we have determined that this is due to a thermal distortion of the resonant coupler that couples the RFD fields to the RFQ. This effect has been mitigated by optimizing the tuning of the resonant coupler during operation. In future designs, the resonant coupler will be stiffer and more intimately coupled to the cooled structures, thus eliminating this thermal distortion effect.

We have learned a lot in the process of solving these problems. These technical problems can be avoided in the future.

## 4 MODIFICATIONS TO REDUCE THE RF POWER REQUIREMENT

Even with solutions to all of these problems, the rf system could not produce enough power to excite the structure to the threshold for proton acceleration. At this point, it would have been expensive to increase the power of the rf system. Instead, we searched for ways to reduce the power requirement of the linac structure.

First, we modified the RFD tank to reduce the design gradient from 7.72 MV/m to 5.90 MV/m. Twelve cells at this reduced gradient resulted in a shorter tank (by 22.86 mm) with reduced drift tube spacing. This required drilling 12 new holes in the keel of the tank for mounting the drift tubes. In order to drill these new holes in virgin metal, the longitudinal plane of the drift tube stems (and RFD lenses) had to be rotated 7.2° about the axis of the tank. This modification reduced the beam energy from 2.50 MeV to 2.00 MeV and the required rf power by 40%. Unfortunately, this did not reduce the required rf power enough to allow operation with the power available from the rf system

Next, we modified the RFD tank to reduce the number of drift tubes from 12 to 9. This shortened the tank by 93.48 mm and reduced the required power by another 19%. This reduced the beam energy of the structure from 2.00 MeV to 1.636 MeV. It was in this configuration that we saw the first accelerated beam.

In summary, the energy and intensity of the beam from the POP was less than expected because of the changes that we had to make to fit within the available rf power. Several design and fabrication flaws raised the rf power requirement above the original estimate, namely, the Q of the cavity (18,600) never got to where it should have been (30,000), the RFD stem design was not optimum, tank modifications left 12 additional drift tube mounting holes to be plugged, the very short tank (less than one diameter) accentuated the end wall and end joint losses, and the very low average beam energy accentuated the drift tube stem losses. Future RFD linac designs will employ a number of improvements that will rectify these problems.

## 5 POTENTIAL USES FOR THE RFD LINAC STRUCTURE

We expect the RFD linac structure to form the basis of a new family of compact, economical, and reliable linac systems serving a whole host of scientific, medical, and industrial applications. The principal medical applications include the production of short-lived radio-isotopes for the positron-based diagnostic procedures (PET and SPECT), the production of epithermal neutron beams for BNCT, and accelerated proton beams for injection into proton synchrotrons to produce the energies required for proton therapy. We also propose an S-Band version of the structure to serve as the 10-70-MeV portion of a 200-MeV booster linac for the proton therapy applications. A modest scientific application includes the production of pulsed cold neutrons for cold neutron physics and related applications[8].

The principal industrial and military applications include the production of intense thermal neutron beams for Thermal Neutron Analysis (TNA), Thermal Neutron Radiography (TNR), and Nondestructive Testing (NDT). High duty factor RFD linac systems could produce nanosecond bursts of fast neutrons to support Pulsed Fast Neutron Analysis (PFNA).

## 6 PLANS FOR CONTINUED DEVELOPMENT OF THE RFD LINAC

Further development of RFD-based linac systems is dependent on further financial support or development contracts. We have mature designs for all components of two different linac systems, which address two different medical applications, namely isotope production for the PET application, and neutron production for the BNCT application. These systems, with minor modification, could be used to satisfy other scientific, medical, and industrial applications.

The PET unit[4], for example, is based on a compact, 12-MeV proton linac with an peak proton beam current of 10 mA and an average proton beam current of 120 µA. This unit, with minor modifications, could be used for production of other isotopes, as injectors for proton synchrotrons, and as injectors for high intensity linear accelerator for energy or materials related applications.

The BNCT unit[5,7], on the other hand, is based on a compact 2.5-MeV proton linac with a peak and average proton beam current of 10 mA. This unit, with minor modifications, could be used for thermal neutron analysis (TNA), neutron activation analysis (NAA), non-destructive testing (NDT), thermal neutron radiography (TNR), explosive detection, and gem irradiation.

The basic principles of the RFD linac structure are now proven, our designs are mature, and we are ready to accept contracts for the development of RFD-based linac systems for practical applications.

## 7 ACKNOWLEDGEMENTS

## REFERENCES

1. D.A. Swenson, "RF-Focused Drift-Tube Linac Structure", LINAC'94, Tsukuba, 1994.
2. D.A. Swenson, Crandall, Guy, Lenz, Ringwall, & Walling, "Development of the RFD Linac Structure", PAC'95, Dallas, 1995.
3. D.A. Swenson, F.W. Guy, K.R. Crandall, "Merits of the RFD Linac Structure for Proton and Light-Ion Acceleration Systems", EPAC'96, Sitges, 1996.
4. D.A. Swenson, K.R. Crandall, F.W. Guy, J.M. Potter, T.A. Topolski, "Prototype of the RFD Linac Structure", LINAC'96, CERN, Geneva, 1996.
5. D.A. Swenson, "CW RFD Linacs for the BNCT Application", CAARI'96, Denton, 1996.
6. D.A. Swenson, K.R. Crandall, F.W. Guy, J.W. Lenz, W.J. Starling, "First Performance of the RFD Linac Structure", LINAC'98, Chicago, 1998.
7. D.A. Swenson, "Compact, Inexpensive, Epithermal Neutron Source for BNCT", CAARI'98, Den., 1998.
8. R.C. Lanza, "Small Accelerator-Based Pulsed Cold Neutron Sources", CAARI'98, Denton, 1998.
9. D.A. Swenson, F.W. Guy, and W.J. Starling, "Commissioning the 2.5-MeV RFD Linac Prototype", PAC'99, New York, 1999.